\documentclass[prl,twocolumn,aps,superscriptaddress,floatfix,10pt]{revtex4}
\usepackage{amsmath}
\usepackage{amssymb}

\usepackage{graphicx}
\begin{document}

\title{Physical Properties of Zener Tunnelling Nano-devices in Graphene}

\author{R. D. Y. Hills}
\affiliation{Department of Physics, Loughborough University, Leicestershire, LE11 3TU, United Kingdom}

\author{F. V. Kusmartsev}
\affiliation{Department of Physics, Loughborough University, Leicestershire, LE11 3TU, United Kingdom}

\begin{abstract}
By considering the direction of charge carriers and the conservation of probablity current the transmission properties of graphene Zener tunnelling nano-devices were obtained. The scattering properties were then used with an adaptation of the Landauer formalism to calculate an analytical expression for current and conductance. The numerical results of the IV characteristics were then briefly discussed for the graphene step and Zener barrier. A comparison between the theoretical model and experimental results shows the similarities of graphene nanoribbons and infinite sheet graphene. This work has been published as \cite{r0}.
\end{abstract}

\maketitle

\pagenumbering{arabic}
\pagestyle{plain}

%%%%%%%%%%
%%%%%%%%%%
%%%%%%%%%%
%%%%%%%%%%
%%%%%%%%%%

\subsection{Introduction}

	Graphene, the thinnest of existing materials is one atom thick and follows a relativistic Dirac electron spectrum. Graphene is very sensitive to its environment, substrate and where and how it is deposited \cite{OHare12}. Charged impurities in a SiO$_2$ substrate induce an electrostatic potential which may confine electron and hole into channels  and puddles as observed in \cite{Martin08a,Martin09b}. The electron doping is usually created by an application of a gate voltage to graphene plane via a metallic strip electrode as in a graphene transistor \cite{Lin11}. It may also be naturally created in epitaxial graphene on SiC due to a terrace step on the SiC layer \cite{Sutter09,Robinson10}. Thus the barrier is modelled by the electrostatic linear potential of the atomic terrace on the SiC substrate or by a gate voltage which scatters relativistic particles, here electrons and holes, moving in the graphene plane.

	It was Klein who showed for relativistic particles that if the potential is on the order of the electron mass, $eV \sim mc^2$, the barrier is nearly transparent (Klein tunnelling) \cite{Klein-1929}. Because of that phenomenon from the first glance the trapping of relativistic particles by potential wells and the existence of bound states within are not possible. On the other hand one may show that a single rectangular potential forming a one dimensional channel extended in other dimension may act similar to the double barrier potential wells \cite{Silvestrov07} which are usually formed in GaAs/AlGaAs \cite{Leo91}. There the resonance bound states may exist, where the potential acts like a tunnelling barrier. The existence of bound states was shown also for a specific rectangular barrier \cite{Pereira06}, which coexist with the Klein tunnelling \cite{Klein-1929}.

%%%%%

	The shape of the potential barrier can alter the tunnelling properties of charge carriers in graphene. For the smooth potentials examined in \cite{c1,d3} the tunnelling characteristics change depending on the particles energy relative to the barrier height. Close to the barrier height the smooth potential shows conventional tunnelling and within the barrier the smooth potential acts as a Fabry-P\'{e}rot interferometer. At energies close to the Dirac point; confined bound states can be found \cite{c1, c2, c3}. The presence of these zero energy bound states is further studied in \cite{c4, c5}, where it is proposed that a top gate can produce an electrostatic potential in graphene in order to model quantum dots.

%%% Zener %%

	The linear dispersion relation that allows for Klein tunnelling through potential barriers also makes graphene an ideal candidate for Zener tunnelling devices. Zener tunnelling is the process whereby an electron may be excited from the valence band into the conduction band by a strong electric field \cite{d1,d2}. In the present paper the Zener tunnelling in graphene nano-devices is represented by an electron-hole interface in the potential structure, which can be seen at energies within a potential step.

%%%%%

	However the linear dispersion relation and Klein tunneling can cause problems for graphene based nanodevices as there is no clear on-off switching. To allow graphene to act in this way an energy gap can be introduced into the energy spectrum. This gap may be formed in zig-zag type nanoribbons \cite{b1} or by interaction with a substate and by transverse electric field \cite{OHare12} in situations when inversion symmetry is broken.

	Graphene nanoribbons (GNRs) that were epitaxially grown on silicon carbide have been shown to act as single channel, room temperature ballistic conductors  \cite{b11}. GNRs with a width of 40-nm were tested using a four point contact method. A 20-nm top gate made from Al$_{2}$O$_{3}$ coated with aluminium allowed the Fermi level of the system to be adjusted. These GNRs showed a large asymmetry with respect to gate voltage caused by np/pn doping and the presence of a semiconducting gap.

	The two-probe measurements of 35 nm wide GNRs; pattered by plasma etching through a PMMA (polymethyl methacrylate) mask on a graphene flake also show a large scale gap, with Fabry-P\'{e}rot resonances arising in the graphene between the contacts and the constriction region when testing for the presence of a quantum dot \cite{b14}.

	The work in \cite{b12} uses back gated lithographically fabricated GNRs with a substrate of highly doped silicon and a 285 nm thick SiO$_2$ gate dielectric. These nanoribbons show a length independent transport gaps with a size inversely proportional to the GNR width and strong non linear IV characteristics when the Fermi energy is within the gap regime.

	GNRs can also be fabricated from mechanically exfoliated graphene sheets on a p-doped Si substrate covered with 300-nm thick SiO$_2$. The GNRs are then formed by oxygen plasma reactive ion etching using a patterned hydrogen silsesquioxane (HSQ) layer as the protective mask \cite{b13}. The conductance through these GNRs resembles that of bulk graphene and an energy gap of approximately 42 meV for one such device.
	
	%%%%%%%%%%%% Topological insulators
	The results obtained in the present paper are equally applicable to a broad class of materials generally named as topological insulators \cite{d4,d5}.  In \cite{d4} it was shown that on the interface between two insulating semiconductors CdTe and HgTe(Se) having inverted band structure there may arise a metallic conducting layer associated with Dirac gapless spectrum;
	there the  single Dirac point %arising at this interface 
	is protected by a time reversal symmetry and the conductivity in the Dirac point at zero temperature tends to infinity.
	%The first real example
	%of topological insulator has been described

	The full energy spectrum of the hexagonal graphene lattice can be obtained from a tight binding approximation. From this two non-equivalent Dirac point {\bf K} and {\bf K'} can be found. A Taylor expansion centered at these points can then produce the Dirac-like Hamiltonians $H_{K}=v_{f}\vec{\sigma} \cdot \hat{p}$ and $H_{K'}=v_{f}\left(\sigma_{x}\hat{p}_{x}-\sigma_{y}\hat{p}_{y}\right)$ \cite{b1}. Together these Hamiltonians can combine to reproduce the 4x4 Hamiltonian from Dirac gamma matrices or inverted band structure heterojunctions \cite{d4}.  Due to the similarities of the two Hamiltonians, here we will only consider the ${\bf K}$ point. An energy gap can be introduced into this Hamiltonian to change the linear energy spectrum of a Dirac point into a parabolic spectrum. The graphene Hamiltonian at a ${\bf K}$ point with an energy gap and an external potential becomes \cite{b2,b3,b4}:
\begin{equation}
	\hat{H}=v_F (\boldsymbol{\sigma}\cdot\hat{p})+\sigma_{z}M\left(x\right)+IV\left(x\right)
	\label{hamiltonian} 	
\end{equation}

	Where $v_{F}$ is the Fermi velocity, $\sigma$ is the Pauli matrices, $\hat{p}$ is the momentum operator, $M(x)$ is an energy gap, $I$ is the identity matrix and $V(x)$ is an exteral potential. The eigenvalues of this system are:
\begin{equation}
	E=V\pm\sqrt{v_{f}^{2}\hbar^{2}\left(k_{x}^{2}+k_{y}^{2}\right)+m^{2}}
	\label{eigenvalue} 
\end{equation}

	Where the values $k_{x}$ and $k_{y}$ are the eigenvalues of the corresponding momentum operator and $m, V$ are constants associated with an external potential $V$ and an energy gap $m$. These eigenvalues have the eigenvector:
\begin{equation}
	\Psi=
	e^{ik_{y}y}
	\left[\begin{array}{ccc}
		e^{iqx}\\
		\alpha e^{iqx+i\theta}
	\end{array}\right]
	\label{psi}
\end{equation}

	Constants have been grouped where possible and are stated as:
\begin{align}
	\alpha&=\frac{\sqrt{\left(E-V\right)^{2}-m^{2}}}{E-V+m}\\
	q^{2}&=\frac{\left(E-V\right)^{2}-m^{2}}{\hbar^{2}v_{f}^{2}}-k_{y}^{2}\\
	\theta&=arctan\left(\frac{k_{y}}{q}\right)
\end{align}

	These wave-functions can then be reduced to the wavefunctions in \cite{b5} by setting $V$ or $m$ to zero as required.
%%%%%%%%%%
%%%%%%%%%%
%%%%%%%%%%
%%%%%%%%%%
%%%%%%%%%%

\subsection{Transmission Properties of Massive Dirac Fermions Through a Potential Step}

\begin{figure}
	\includegraphics[scale=0.2]{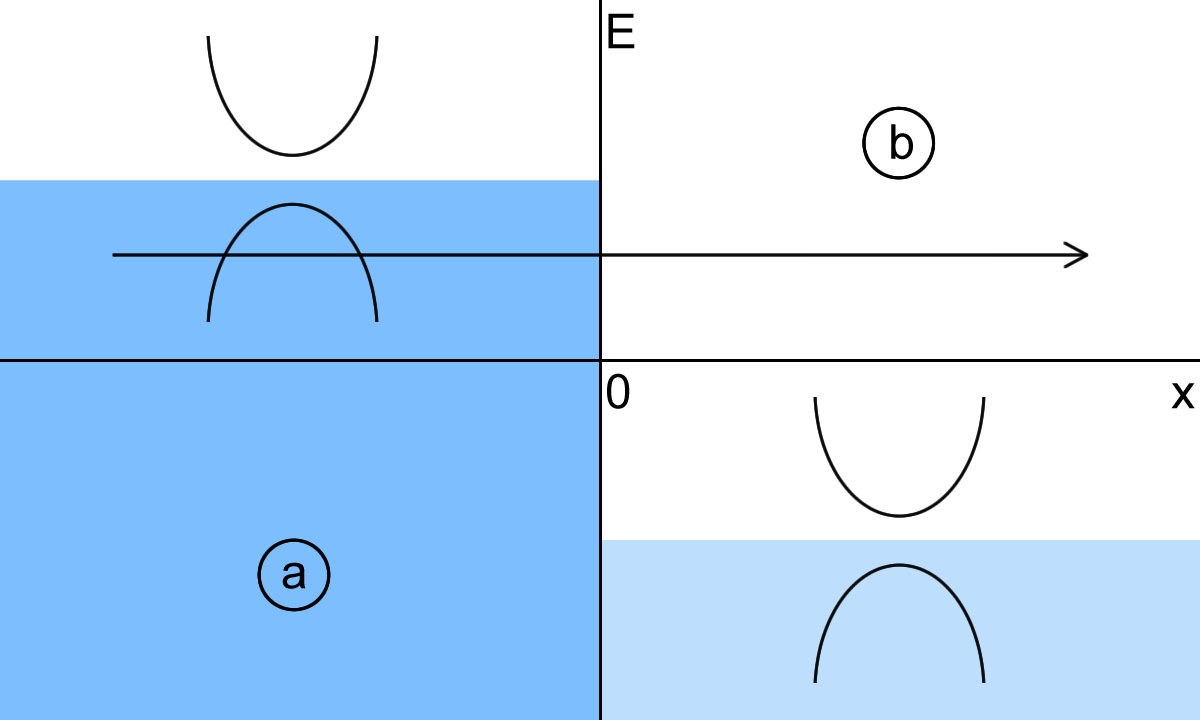}
	\caption{(Colour Online) Diagram of a massive potential step in graphene, including the energy spectrum in specific regions. Here a right traveling massive charge carrier is shown transmitting through a symmetrical step with heights $V_{a}=V_{0}$ and $V_{b}=-V_{0}$, representing Zener tunnelling.}
	\label{rectangular-step-flat}
\end{figure}

	The massive graphene potential step is a 2-region system shown in FIG.\ref{rectangular-step-flat}.	To find the transmission properties of this system, the wave-functions from eq.(\ref{psi}) can be used to describe the left and right of the step interface. By requiring continuity at this interface a system of simultaneous equations can be created.
\begin{equation}
	e^{ik_{y}y}\left(e^{iq_{a}x}+re^{-iq_{a}x}\right)=t_{s}e^{iq_{b}x}e^{ik_{y}y}
	\label{simultaneous equations 1}
\end{equation}
\begin{equation}
	e^{ik_{y}y}\left(\alpha_{a}e^{iq_{a}x+i\theta_{a}}-r\alpha_{a}e^{-iq_{a}x-i\theta_{a}}\right)=t_{s}\alpha_{b}e^{iq_{b}x+i\theta_{b}}e^{ik_{y}y}
	\label{simultaneous equations}
\end{equation}

	The subscripts $a,b$ have been included to represent the potentials and energy gaps in the regions shown in FIG.\ref{rectangular-step-flat}. Solving these simultaneous equations at the barrier interface $x=0$ for $t_{s}$ produces:
\begin{equation}
	t_{s}=\frac{2\alpha_{a}cos\left(\theta_{a}\right)}{\alpha_{a}e^{-i\theta_{a}}+\alpha_{b}e^{i\theta_{b}}}
\end{equation}

	However, if instead of solving this set of equations, a transfer matrix method is used \cite{b18} with left and right travelling waves from eq.(\ref{psi}) the system becomes:
		\begin{align}
			\left[\begin{array}{ccc}
				e^{iq_{a}x}&e^{-iq_{a}x}\\
				\alpha_{a}e^{iq_{a}x+i\theta_{a}}&-\alpha_{a}e^{-iq_{a}x-i\theta_{a}}
			\end{array}\right]
			\left[\begin{array}{ccc}
				a_{1}\\
				a_{2}
			\end{array}\right]
			\hspace{1cm}
			\\=
			\left[\begin{array}{ccc}
				e^{iq_{b}x}&e^{-iq_{b}x}\\
				\alpha_{b}e^{iq_{b}x+i\theta_{b}}&-\alpha_{b}e^{-iq_{b}x-i\theta_{b}}
			\end{array}\right]
			\left[\begin{array}{ccc}
				a_{3}\\
				a_{4}
			\end{array}\right]
		\end{align}
Setting the step interface at $x=0$, the transmission coefficient $t_{m}$ can then be evaluated to:
\begin{equation}
	t_{m}=\frac{2\alpha_{b}e^{-i\theta_{b}}cos\theta_{b}}{\alpha_{b}e^{2i\theta_{b}}+\alpha_{a}e^{-i\theta_{a}+i\theta_{b}}}
\end{equation}

	This discrepency in $t_{s}$ and $t_{m}$ is due to the transmission probability not being defined as $|t|^2$. As discussed in \cite{b4, d3} the conservation of probability current density can provide the correct method for calculating the transmission through the system. The conservation of current is defined as:
		\begin{equation}
			\frac{d}{dt}|\psi|^{2}+\nabla\cdot {\bf j}=0
		\end{equation}
As the system here is time independent only the probability current;
		\begin{equation}
			{\bf j} =\psi^{*} {\bf \sigma} \psi
		\end{equation}
needs to be considered. From the continuity equation; the probability current into the system must equal the probability current out of the system.
		\begin{align}
			j_{i}=j_{t}+j_{r}
			\hspace{1cm}
			1=\frac{j_{t}}{j_{i}}+\frac{j_{r}}{j_{i}}
			\label{current density}
		\end{align}
Using the graphene wave-functions in one dimension from the left and right of the step interface in eq.(\ref{simultaneous equations 1}) and eq.(\ref{simultaneous equations}) with eq.(\ref{current density}) shows that the transmission $T$ is in fact:
\begin{equation}
	T=|t|^{2}\frac{\alpha_{b}cos\left(\theta_{b}\right)}{\alpha_{a}cos\left(\theta_{a}\right)}
	\label{ttT}
\end{equation}

	The relation in eq.(\ref{ttT}), when used with $t_{s}$ then produces the correct result for the transmission $T$. Due to the extra left travelling wave in region $b$ required to construct a transfer matrix; the result for $t_{m}$ does not agree with this result and cannot be use to calculate the transmission. However, by using the relation $T=1-R$ and the reflection coefficient the transfer matrix method produces a result consistent with the simultaneous equations method. Due to the increasing complexity of using the simultanious equations method to solve barrier systems the transfer matrix method will be used for larger systems, but only via the relation $T=1-R$. By using $T=1-R$ with the transfer matrix method or equation (6) with simultaneous equations; the result for transmission through a graphene step is:
\begin{equation}
	T=\frac{4\alpha_{a}\alpha_{b}cos\left(\theta_{a}\right)cos\left(\theta_{b}\right)}{\alpha_{a}^{2}+\alpha_{b}^{2}+2\alpha_{a}\alpha_{b}cos\left(\theta_{a}+\theta_{b}\right)}
	\label{stept}
\end{equation}

\begin{figure}
	\includegraphics[scale=0.2]{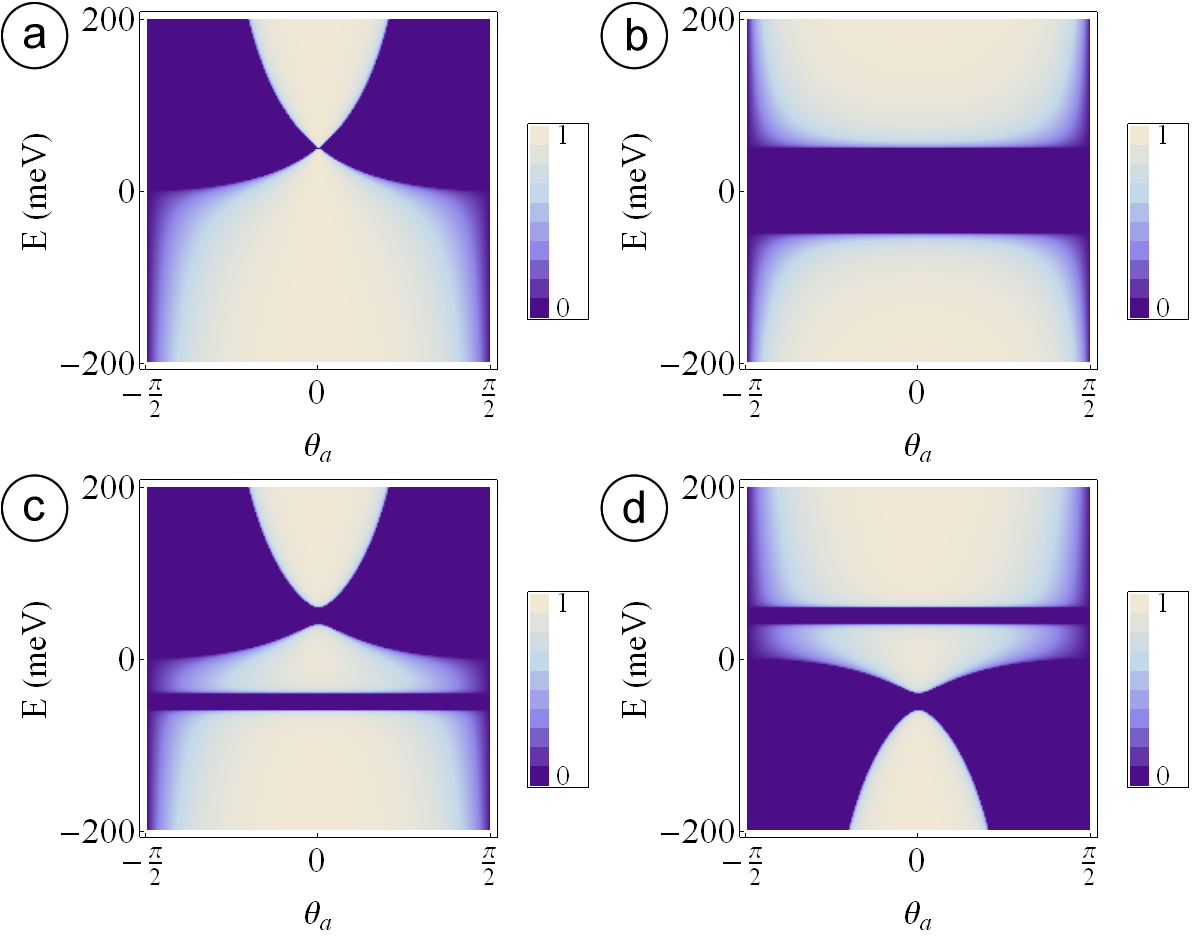}
	\caption{(Colour Online) Density plots of transmission against energy and incident angle for various two regions systems. (a) Potential step with $V_{a}=50$ meV, $V_{b}=-50$ meV. (b) Massive step with $m_{a}=50$ meV and $m_{b}=0$ meV. (c) Massive potential step where $V_{a}=50$ meV, $V_{b}=-50$ meV and $m_{a,b}=10$ meV. (c) Opposite massive potential step where $V_{a}=-50$ meV, $V_{b}=50$ meV and $m_{a,b}=10$ meV.}
	\label{step-t-flat}
\end{figure}

	To obtain the results shown in FIG.\ref{step-t-flat} the correct direction of the charge carriers must be considered. At energies within the step, i.e. $|E|<|V|$ there will be an electron-hole interface. Due to their opposite charge a charge-carrying hole will need to move in the opposite direction to an electron \cite{d3}. This difference in direction is represented by a change in incident angle. The incident angle for holes is therefore given as $\theta_{h}=\pi-\theta_{e}$. Using this phase shift the transmission probability will never exceed one.

	The result in eq.(\ref{stept}) can be used for a variety of two region systems. By removing the corresponding terms potential steps, massive steps and the combination of both can be plotted. Examples of these systems can be seen in FIG.\ref{step-t-flat}. When the anglular dependence in eq.(\ref{stept}) is removed the transmission probability becomes equal to one at all energies (except $E=V$), as seen in the one dimensional Klein tunneling case in \cite{d3}. The plot in FIG.\ref{step-t-flat}(a) shows strong agreement to the semi-classical result for the graphene potential step also featured in \cite{d3}.
%%%%%%%%%%
%%%%%%%%%%
%%%%%%%%%%
%%%%%%%%%%
%%%%%%%%%%

\subsection{Transmission Properties of Massive Dirac Fermions Through a Zener Tunnelling Barrier}
\begin{figure}
	\includegraphics[scale=0.2]{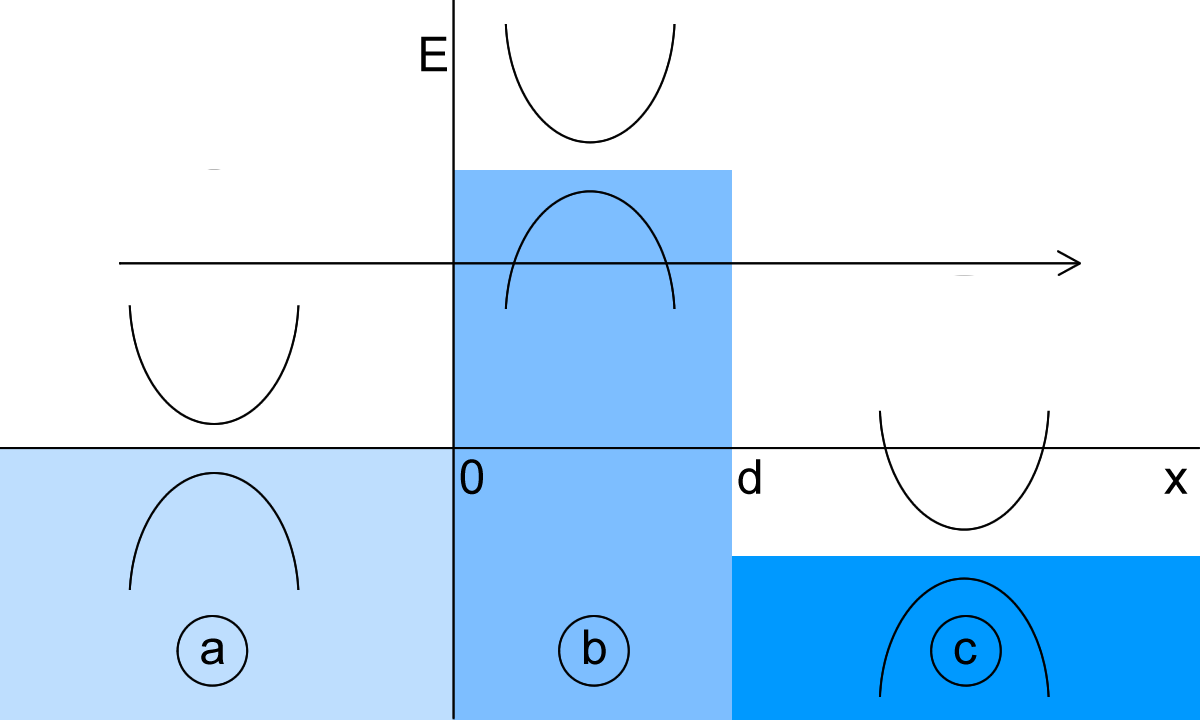}
	\caption{(Colour Online) Diagram of a Zener tunnelling barrier in graphene, including energy spectrum in specific regions. Here a right travelling massive charge carrier is shown transmitting through a barrier with width $d$ and heights $V_{b}>V_{a}$ and $V_{c}<V_{a}$.}
	\label{asy-flat}
\end{figure}

	The massive Zener tunnelling potential barrier is a three-region system similar to the potential barrier with the exception that regions $a$ and $c$ are not equal. This three-region system can be interpreted as a double potential step, or as a Zener tunnelling barrier; which includes a barrier on top of a step shown in FIG.\ref{asy-flat}.
\begin{figure}
	\includegraphics[scale=0.2]{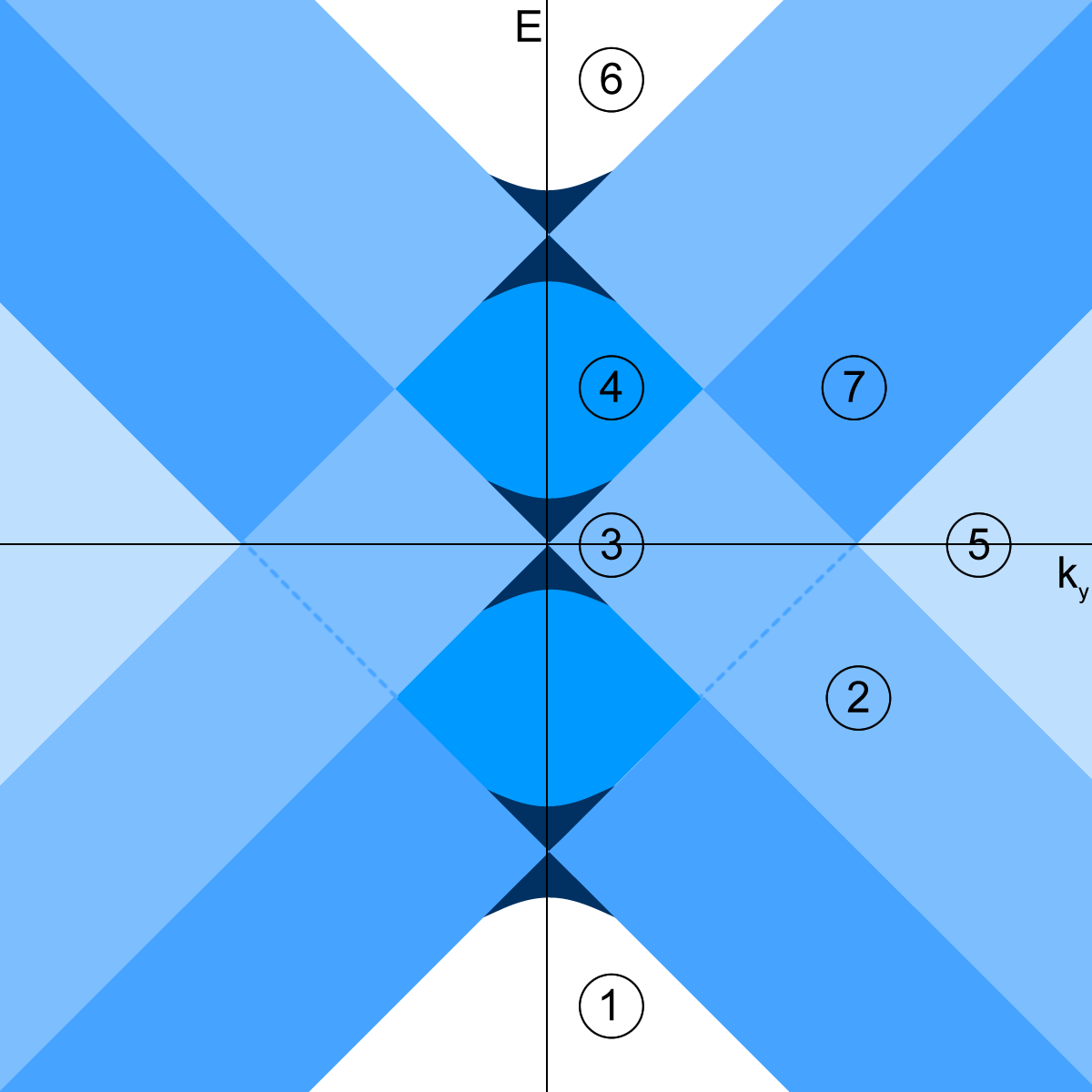}
	\caption{(Colour Online) Transmission region diagram of energy against y-momentum ($\hbar v_{f}=1$) for a three region Zener tunnelling system. Regions of transport (1, 4, 6), energy gap (3) and no propagation (2, 5, 7) are shown.}
	\label{double-step-regions-flat}
\end{figure}

	The transmission properties of the Zener tunnelling barrier, or "Zener barrier" is shown in FIG.\ref{double-step-regions-flat}. In the same way as a symmetrical barrier; regions 1, 4 and 6 are where electron-electron, electron-hole-electron, electron-hole or hole-hole transport occur, here it is expected that tranmission is high as well as evidence of resonances and bound states. In region 3 the energy gap introduced into the graphene spectrum causes no transmission or no incident particles depending on the direction of the incident charge carrier. Region 5 is only represented on this diagram, $\hbar v_{f} k_{y}>E$ here, only imaginary solutions could exist here. Regions 2 and 7 are regions where $\hbar v_{f} k_{y}>E$ but  $\hbar v_{f} k_{y}<E-V$, therefore there is no transmission here, but bound states can still be found for these regions.

	To find the scattering properties for a Zener barrier the transfer matrix method \cite{b18} outlined in the previous section can be used. Using eq.(\ref{psi}) and introducing left and right travelling waves, the wave-functions in each region are:
\begin{align}
	\psi_{a}&=e^{ik_{y}y}
	\left[\begin{array}{ccc}
		e^{iq_{a}x}&e^{-iq_{a}x}\\
		\alpha_{a}e^{iq_{a}x+i\theta_{a}}&-\alpha_{a}e^{-iq_{a}x-i\theta_{a}}
	\end{array}\right]
	\left[\begin{array}{ccc}
		a_{1}\\
		a_{2}
	\end{array}\right]\\
	\psi_{b}&=e^{ik_{y}y}
	\left[\begin{array}{ccc}
		e^{iq_{b}x}&e^{-iq_{b}x}\\
		\alpha_{b}e^{iq_{a}x+i\theta_{b}}&-\alpha_{b}e^{-iq_{b}x-i\theta_{b}}
	\end{array}\right]
	\left[\begin{array}{ccc}
		a_{3}\\
		a_{4}
	\end{array}\right]\\
	\psi_{c}&=e^{ik_{y}y}
	\left[\begin{array}{ccc}
		e^{iq_{c}x}&e^{-iq_{c}x}\\
		\alpha_{c}e^{iq_{c}x+i\theta_{c}}&-\alpha_{c}e^{-iq_{c}x-i\theta_{c}}
	\end{array}\right]
	\left[\begin{array}{ccc}
		a_{5}\\
		a_{6}
	\end{array}\right]
\end{align}

\begin{figure}
	\includegraphics[scale=0.2]{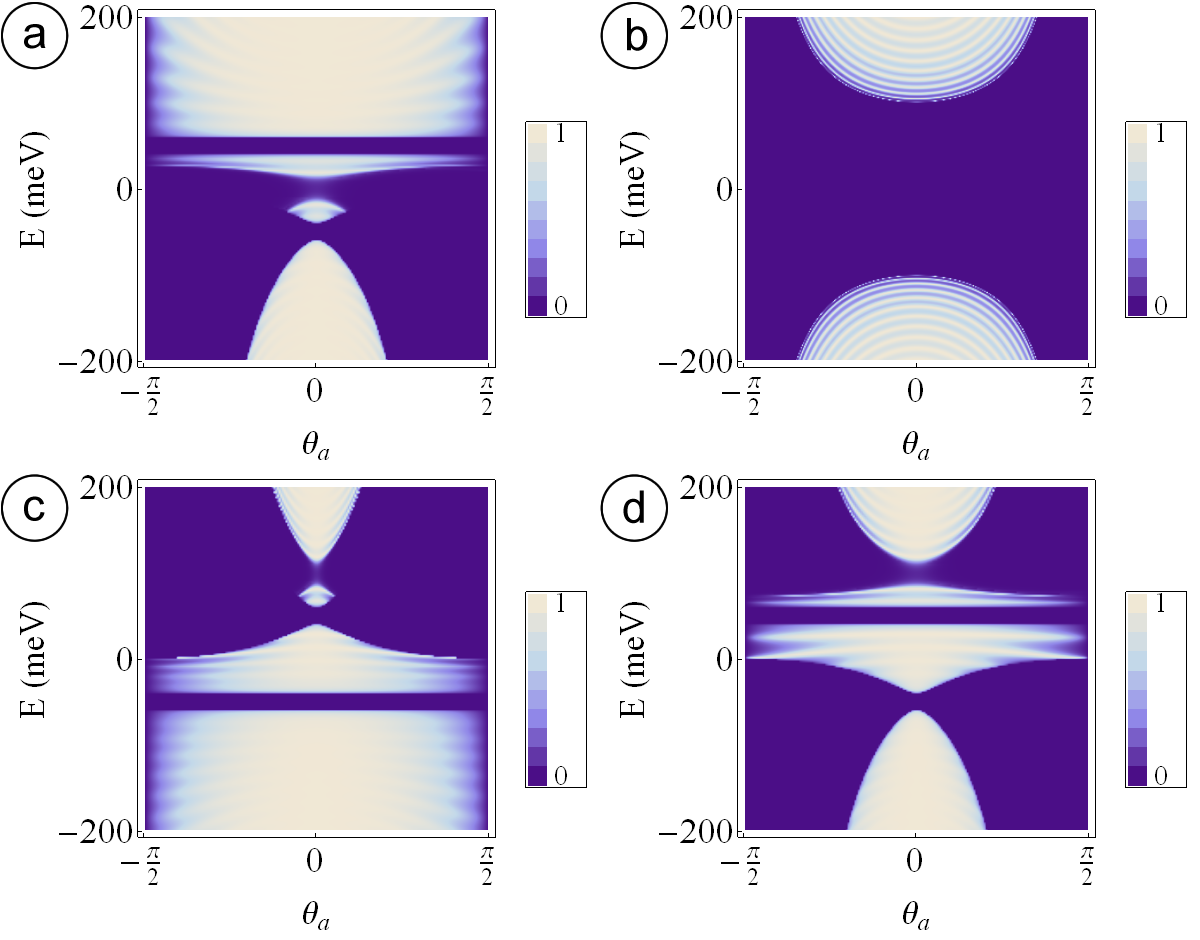}
	\caption{(Colour Online) Density plot of transmission with energy against incident angle for a three-region system. For all plots $d=200$ nm. (a) The double step where $V_{a}=50$ meV, $V_{b}=0$, $V_{c}=-V_{a}$ and $m_{a,b,c}=10$ meV. (b) A massive Zener barrier with $m_{a}=50$ meV, $m_{b}=100$ meV and $m_{c}=0$ meV. (c) A Zener barrier with $V_{a}=-50$ meV, $V_{b}=100$ meV and $V_{c}=50$ meV. (d) A Zener barrier with $V_{a}=50$ meV, $V_{b}=100$ meV and $V_{c}=-50$ meV.}
	\label{asy-t-flat}
\end{figure}

Where regional subscripts have been added to identify the wave-functions in each region. The continuity of these wave-functions at the barrier interfaces requires that at $x=0$, $\psi_{a}=\psi_{b}$, which in matrix form:
		\begin{align}
			\left[\begin{array}{ccc}
				1&1\\
				\alpha_{a}e^{i\theta_{a}}&-\alpha_{a}e^{-i\theta_{a}}
			\end{array}\right]
			\left[\begin{array}{ccc}
				a_{1}\\
				a_{2}
			\end{array}\right]
			&=
			\left[\begin{array}{ccc}
				1&1\\
				\alpha_{b}e^{i\theta_{b}}&-\alpha_{b}e^{-i\theta_{b}}
			\end{array}\right]
			\left[\begin{array}{ccc}
				a_{3}\\
				a_{4}
			\end{array}\right]
		\end{align}
		For convenience these matrices will be labeled $m_{1,2}$ so that:
		\begin{align}
			m_{1}\left[\begin{array}{ccc}
				a_{1}\\
				a_{2}
			\end{array}\right]
			&=
			m_{2}\left[\begin{array}{ccc}
				a_{3}\\
				a_{4}
			\end{array}\right]
		\end{align}
and at $x=d$, $\psi_{b}=\psi_{c}$, the wave-functions can be expressed as:
		\begin{align}
			\left[\begin{array}{ccc}
				e^{iq_{b}d}&e^{-iq_{b}d}\\
				\alpha_{b}e^{iq_{b}d+i\theta_{b}}&-\alpha_{b}e^{-iq_{b}d-i\theta_{b}}
			\end{array}\right]
			\left[\begin{array}{ccc}
				a_{3}\\
				a_{4}
			\end{array}\right]
			\hspace{1cm}\\
			=
			\left[\begin{array}{ccc}
				e^{iq_{c}d}&e^{-iq_{c}d}\\
				\alpha_{c}e^{iq_{c}d+i\theta_{c}}&-\alpha_{c}e^{-iq_{c}d-i\theta_{c}}
			\end{array}\right]
			\left[\begin{array}{ccc}
				a_{5}\\
				a_{6}
			\end{array}\right]
		\end{align}
		and the matrices will be labeled $m_{3,4}$ so that:
		\begin{align}
			m_{3}\left[\begin{array}{ccc}
				a_{3}\\
				a_{4}
			\end{array}\right]
			=
			m_{4}\left[\begin{array}{ccc}
				a_{5}\\
				a_{6}
			\end{array}\right]
		\end{align}
		These definitions allow the system to be solved for the incident and transmitted coefficients and produce the transfer matrix $M$:
		\begin{align}
			\left[\begin{array}{ccc}
				a_{5}\\
				a_{6}
			\end{array}\right]=M
			\left[\begin{array}{ccc}
				a_{1}\\
				a_{2}
			\end{array}\right]
		\end{align}
where:
\begin{equation}
	M=m_{4}^{-1}m_{3}m_{2}^{-1}m_{1}
\end{equation}
From this transfer matrix and the relation $T=1-R$ the transmission through the structure becomes:
\begin{equation}
	T=\frac{t_{1}}{t_{2}t_{3}}
	\label{asy-t}
\end{equation}
\begin{align}
t_{1}&=4\alpha_{a}\alpha_{b}^{2}\alpha_{c}cos(\theta_{a})cos(\theta_{c})cos^{2}(\theta_{b})e^{i\theta_{a}+i\theta_{c}}
\\	t_{2}&=\alpha_{b}\alpha_{c}C_{-}+\alpha_{a}\alpha_{b}e^{i\theta_{a}+i\theta_{c}}C_{+}+i\left(\alpha_{a}\alpha_{c}e^{i\theta_{a}}+\alpha_{b}^{2}e^{i\theta_{c}}\right)S
\\
t_{3}&=\alpha_{a}\alpha_{b}C_{+}+\alpha_{b}\alpha_{c}e^{i\theta_{a}+i\theta_{c}}C_{-}-i\left(\alpha_{b}^{2}e^{i\theta_{a}}+\alpha_{a}\alpha_{c}e^{i\theta_{c}}\right)S
\end{align}
Where $C_{\pm}=cos(dq_{b}\pm\theta_{b})$ and $S=sin(dq_{b})$. By considering the correct incident angles for inside the step and barrier regions the plots in FIG.\ref{asy-t-flat} were obtained.

	The result in eq.(\ref{asy-t}) can be varified by reducing it to previously obtained results. If the constants in region $a$ and $c$ are equivalent so that $\alpha_{a}=\alpha_{c}$ and $\theta_{a}=\theta_{c}$, this result will become the transmission through the potential barrier shown in \cite{b1}. Similarly if $\alpha_{b}=\alpha_{c}$, $\theta_{b}=\theta_{c}$ and $d=0$ the system will be reduced to a potential step and eq.(\ref{asy-t}) will be equal to eq.(\ref{stept}) as expected.

	The extra boundary at $x=d$ causes an additional reflected term into region $b$. This extra term allows the three-region system to act as a Fabry-P\'{e}rot resonator. Under the resonance condition for a potential barrier $dq_{b}=n\pi$ \cite{b6} an expression for resonances inside the barrier can be obtained:
\begin{equation}
	E=V_{b}\pm\sqrt{\hbar^{2}v_{f}^{2}\left(\frac{n^{2}\pi^{2}}{d^{2}}+k_{y}^{2}\right)+m_{b}^{2}}
	\label{resonances}
\end{equation}

	When the resonance condition is applied to eq.(\ref{asy-t}) the expression simplifies to:
\begin{equation}
	T=\frac{4\alpha_{a}\alpha_{c}cos\left(\theta_{a}\right)cos\left(\theta_{c}\right)}{\alpha_{a}^{2}+\alpha_{c}^{2}+2\alpha_{a}\alpha_{c}cos\left(\theta_{a}+\theta_{c}\right)}
	\label{}
\end{equation}
	
	This result is identical to eq.(\ref{stept}); under resonance conditions the barrier becomes transparent, leaving only the step produced between regions $a$ and $c$ to scatter charge carriers. 

	The step-like properties re-occur when examining large potentials required for Klein tunnelling. With large potentials $\theta_{b}=0$, $V_{b}>>E$, $\alpha_{b}=-1$ the transmission can be reduced to show step-like transmission properties with the addition of non-theta dependent resonances. To varify this the resonance condition can then be applied with $V_{b}>>k_{y},m$ to find resonances close to the Fermi level:
	\begin{equation}
		E=\pm \frac{\hbar v_{f}n \pi}{d}
		\label{klein-res}
	\end{equation}
	With the resonance condition this result further reduces to the transmission for the potential step.

	As shown in \cite{b7} a single potential barrier in graphene should show signs of bound states within the barrier. By requiring growth-decay wave-functions as eigenvectors to the Hamiltonian in eq.(\ref{hamiltonian}) a system of growth-oscillatory-decay can be created. Growth-decay wave-functions take the form of:
\begin{equation}
	\psi_{gd}=
	e^{ik_{y}y}
	\left[\begin{array}{ccc}
		e^{q_{d}x}+e^{-q_{d}x}\\
		i\alpha_{-}e^{q_{d}x}-i\alpha_{+}e^{-q_{d}x}
	\end{array}\right]
	\label{psigd}
\end{equation}
where the constants $q_{d}$ and $\alpha_{\pm}$ have been grouped together and are defined as:
\begin{align}
	q_{d}^{2}&=k_{y}^{2}-\frac{\left(E-V\right)^{2}+m^{2}}{\hbar^{2}v_{f}^{2}}\\
	\alpha_{\pm}&=\frac{\hbar v_{f}}{V-E-m}\left(q_{d}\pm k_{y}\right)
\end{align}

When combined with the wave-functions in eq.(\ref{psi}) the bound states can then be found from the solutions of the equation:
\begin{equation}
	tan\left(dq_{b}\right)=-\frac{\varepsilon q\left(\alpha_{+}-\alpha_{-}\right)}{-\alpha_{-}\alpha_{+}+\alpha_{b}\alpha_{b}^{*}+\varepsilon k_{y}\left(\alpha_{+}+\alpha_{-}\right)}
	\label{boundstates}
\end{equation}
			where $\varepsilon =\hbar v_{f}/\left(E-V_{b}\right)$. With the exception of an energy gap at $V_{b}$, this equation produces results very similar results to the previously published result. The Zener tunnelling region introduced below the system has little effect on the bound states found within the barrier due to solutions decaying exponentially away from the barrier.

%%%%%%%%%%
%%%%%%%%%%
%%%%%%%%%%
%%%%%%%%%%
%%%%%%%%%%

\subsection{IV Characteristics}
	The current through the scattering systems formulated earlier can be calculated with the Landauer formalism for ballistic transport. In this model perfect electron emitters are connected to a scattering device via perfectly conducting wires. The electron emitters emit electrons up to the quasi-Fermi-energy $\mu_{L}$ and $\mu_{R}$ into the respective side of the scattering device. In this model the current through the scattering device is given in \cite{b8} as:
\begin{equation}
	I=ev_{f}\frac{dn}{dE}T\left(\mu_{L}-\mu_{R}\right)
	\label{current-dos}
\end{equation}
where $e$ is the electron charge, $v_{f}$ is the Fermi velocity and $dn/dE$ is the density of states. At a finite temperature the electron emitters inject electrons as described by the Fermi-Dirac distribution:
\begin{equation}
	f_{L,R}=f\left(E-\mu_{L,R}\right)=\frac{1}{e^{\frac{E-\mu_{L,R}}{k_{b}t}}+1}
\end{equation}
instead of up to the quasi-Fermi-energies $\mu_{L}$ and $\mu_{R}$. Here $k_{b}$ is the Boltzman constant and $t$ is the temperature. Using the density of states for graphene in \cite{b1}:
\begin{equation}
	\frac{dn}{dE}=\frac{2A_{c}}{\pi}\frac{|E|}{v_{f}^{2}}
	\label{graphene-dos}
\end{equation}
where $A_{c}=3\sqrt{3}a^{2}/2$ is the area of the unit cell and $a$ is the carbon-carbon distance. Integrating over energy and incident angle produces the $x$-direction current:
\begin{equation}
	I_{x}=I_{0}\int^{\infty}_{-\infty}\int^{\pi/2}_{-\pi/2}T\left(E,\theta\right)\left[f_{L}-f_{R}\right]|E|cos\left(\theta\right)dEd\theta
	\label{i}
\end{equation}
with the constant $I_{0}=e\frac{2L_{y}}{\pi\hbar^{2}v_{f}}$ and $L_{y}$ is the length of the system in the y-direction. At this stage the current shows a similar form to that in \cite{b15, b16}, with the exception that the graphene density of states causes an additional $|E|$ term and the graphene transmission probability introduces a theta dependence.

\begin{figure}
	\includegraphics[scale=0.2]{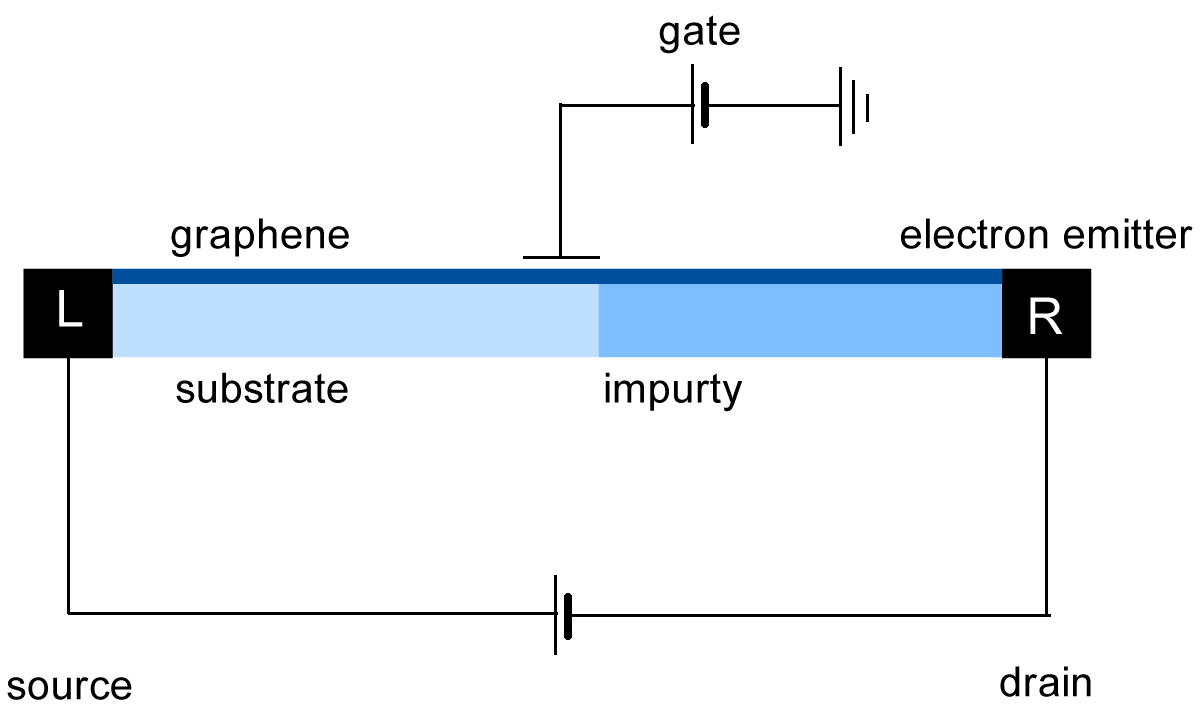}
	\caption{(Colour Online) A simple example of a graphene transistor. A second substrate to the right of the gate region creates Zener tunnelling properties.}
	\label{transistor-flat}
\end{figure}

	The conductance of a device is $G=I/V$. From the Landauer formalism the voltage accross the system is given by $eV=R\left(\mu_{L}-\mu_{R}\right)$, where $R$ is the reflection probability. This expression must then be adjusted for non-zero temperatures and converted to the quasi-Fermi-energies of the electron emitters.

	This definition of voltage provides an interesting property for graphene systems; the transmission through graphene devices can become one due to Klein tunneling or resonance conditions causing the reflection to become zero, resulting in zero voltage. To provide a logical result for conductance a method must be derived that allows for zero reflection. As stated in \cite{b8} a system with many scattering devices will cause any incident electrons to eventually be scattered so that $R \approx 1$. This system is representitive of a single scattering device with multiple voltage or current probes. Alternatively, it is suggested in \cite{b9} that any electric field can be absorbed by a finite region of perfect conductor, again resulting in the removal of the reflection probability from the conductance calculation.

	Using this expression for voltage, with $R \approx 1$,  the methods in \cite{b8} and the graphene density of states, the conductance at a finite temperature is:
\begin{equation}
	G_{x}=G_{0}\int^{\pi/2}_{-\pi/2} \int^{\infty}_{-\infty} T\left(E,\theta\right)\frac{f_{L}-f_{R}}{\mu_{L}-\mu_{R}}|E|cos\theta dE d\theta
\end{equation}
with $G_{0}=e^{2}\frac{2L_{y}}{\pi\hbar^{2}v_{f}}$. However as this result is very similar to the definition of current, the conductance will be considered at zero temperature and for small voltages. At zero temperature and for small voltages, the Fermi distributions become the Dirac delta function centered at the Fermi energy $E_{f}$. With the identity $\int f(x)\delta(x) dx=f(0)$ the zero temperature conductance for small voltages becomes:
\begin{equation}
	G_{x}= G_{0}\int^{\frac{\pi}{2}}_{-\frac{\pi}{2}} T\left(E_{f},\theta\right)|E_{f}|cos\left(\theta\right)d\theta
	\label{g}
\end{equation}

	This result for conductance includes the Fermi energy, as required from the density of states of graphene and the integration of a Dirac delta function. The full derivation of the expressions in this section can be found in the supplementary information \cite{b17}. A similar result for conductance is shown in \cite{b6,b19}, however many published expressions for conductance do not include this term \cite{b2,b4,b10}. The inclusion of the Fermi energy causes the conductance to become linear outside of the step, or barrier region, dramatically changing the result obtained.

\begin{figure}
	\includegraphics[scale=0.2]{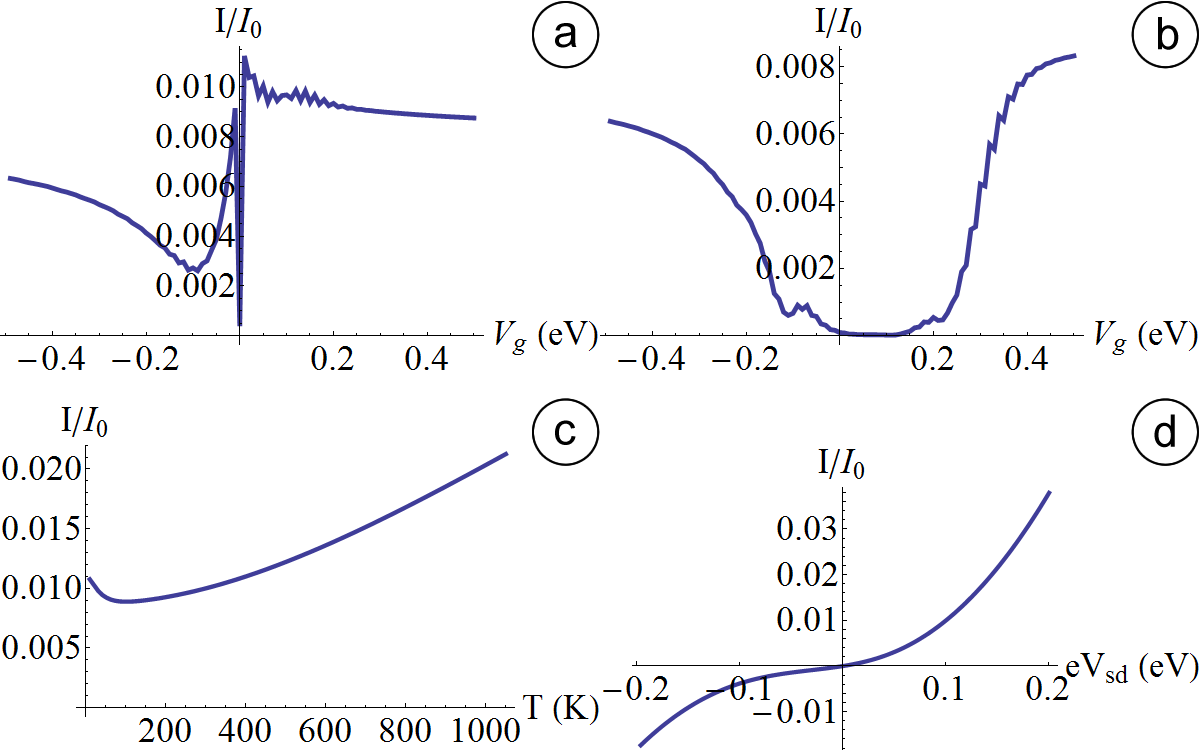}
	\caption{Numerical IV characteristics for graphene steps computed from eq.(\ref{i}) with $I_{0}=e2L_{y}/\pi\hbar^{2}v_{f}$. (a) Current against step height $V_{g}$ where $V_{a}=V_{g}$, $V_{b}=-V_{g}$, $eV_{sd}=100$ meV and $t=298$ K. (b) Current against step height as in (a) with $m_{a}=200$ meV. (c) Current against temperature with $V_{a}=50$ meV, $V_{b}=-50$ meV and $eV_{sd}=100$ meV. (d) Current against $eV_{sd}$ where $V_{a}=50$ meV, $V_{b}=-50$ meV and $t=298$ K.}
	\label{step-i-flat}
\end{figure}

\begin{figure}
	\includegraphics[scale=0.2]{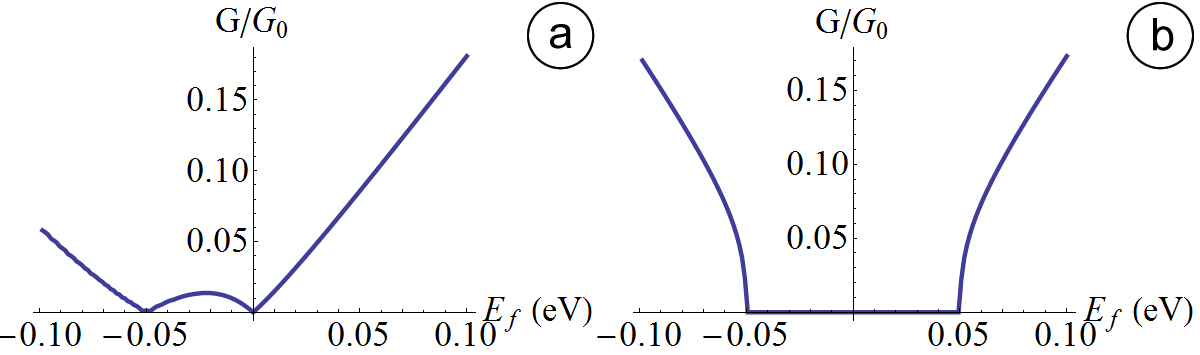}
	\caption{Numerical conductance plots for graphene steps from eq.(\ref{g}) with $G_{0}=e^{2}2L_{y}/\pi\hbar^{2}v_{f}$. (a) Conductance for a graphene potential step with $V_{a}=50$ meV and $V_{b}=-50$ meV. (b) Conductance for a graphene massive step with $m_{a}=50$ meV and $m_{b}=0$ meV.}
	\label{step-g-flat}
\end{figure}

	The numerical IV characteristics for a graphene potential step with eq.(\ref{i}) and eq.(\ref{g}) are shown in FIG.\ref{step-i-flat} and FIG.\ref{step-g-flat} respectivly where $eV_{sd}$ is the difference in chemical potentials of the two electron emitters $\mu_{L}-\mu_{R}$. 

	The orientation of the step greatly influences the current at low step heights. The plot in FIG.\ref{step-i-flat}(a) shows a symmetrical step with $V_{a}=V_{g}$ and $V_{b}=-V_{g}$. If the step direction were reversed, the results are reversed.

	The plot in FIG.\ref{step-i-flat}(d) shows an IV curve similar to that of a traditional Zener diode. If an energy gap is introduced into this system the region of zero current centered at $eV_{sd}=0$ eV will expand out accross the voltage axis. However, a large energy gap is required in order to mimic Zener diodes; an energy gap of $200$ meV created a zero current region to approximately $100$ mV.

	From FIG.\ref{step-i-flat}(c) a fairly linear temperature dependence can be seen at higher temperatures, showing a larger voltage dependence at lower temperatures.

	The conductance plots in FIG.\ref{step-g-flat} then show a largely linear dependence due to the $|E_{F}|$ term in eq.(\ref{g}). The exceptions here are caused inside the step where the conductance reduces near $E \approx V$ or $|E|<m$.

\begin{figure}
	\includegraphics[scale=0.2]{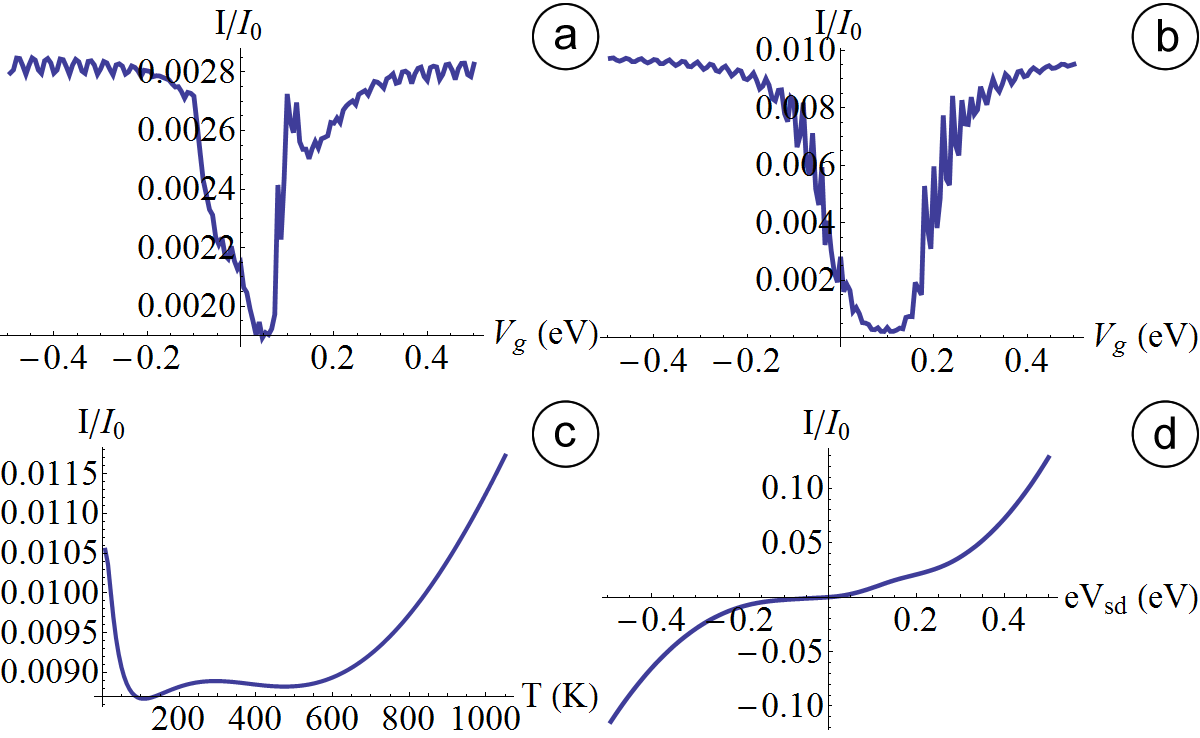}
	\caption{Numerical IV characteristics for three region Zener tunnelling graphene systems computed from eq.(\ref{i}) with $I_{0}=e2L_{y}/\pi\hbar^{2}v_{f}$. In all plots the gate region with subscript $b$ has a width $d=100$ nm. (a) Current against gate voltage where $V_{a}=-100$ meV, $V_{b}=V_{g}$, $V_{c}=100$ meV, $eV_{sd}=100$ meV and $t=298$ K. (b) Current againt gate voltage with an energy gap, $V_{a}=100$ meV, $V_{b}=V_{g}$, $V_{c}=-100$ meV, $m_{a,c}=0$ meV, $m_{b}=100$ meV, $eV_{sd}=100$ meV and $t=298$ K. (c) Current againt temperature where $V_{a}=100$ meV, $V_{b}=200$ meV, $V_{c}=-100$ meV and $eV_{sd}=100$ meV. (d) Current against $eV_{sd}$ as in (c) with $t=298$ K.}
	\label{asy-i-flat}
\end{figure}

\begin{figure}
	\includegraphics[scale=0.2]{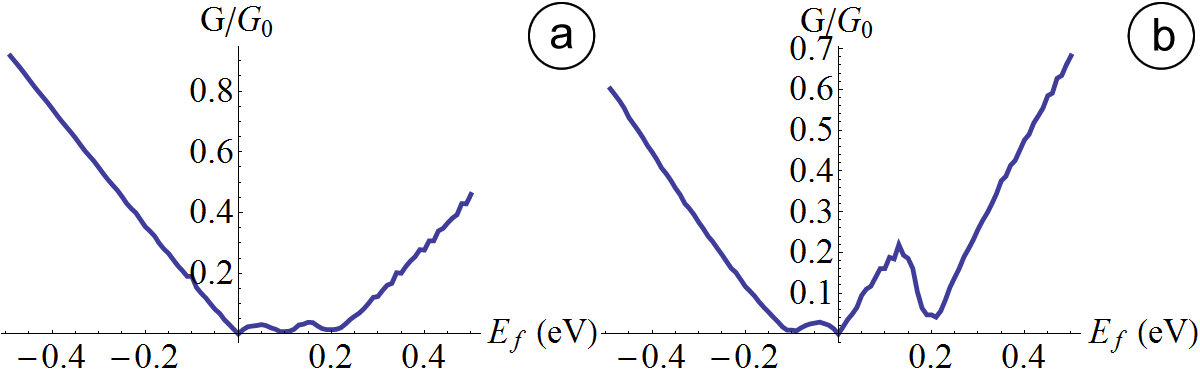}
	\caption{Numerical conductance plots for three region Zener tunnelling graphene systems from eq.(\ref{g}) with $G_{0}=e^{2}2L_{y}/\pi\hbar^{2}v_{f}$ and  width $d=100$ nm. (a) $V_{a}=-100$ meV, $V_{b}=200$ meV and $V_{c}=100$ meV. (b) $V_{a}=100$ meV, $V_{b}=200$ meV and $V_{c}=-100$ meV.}
	\label{asy-g-flat}
\end{figure}

	The IV characteristics and conductance plots for three region systems are then shown in FIG.\ref{asy-i-flat} and FIG.\ref{asy-g-flat}. At $V_{g}=0$ the Zener barrier is reduced to a step and by examining FIG.\ref{asy-i-flat}(a) shows clear step-like current properties when the gate voltage is equal to the step heights. When the gate voltage exceeds the step region clear oscillations are visible caused by the second barrier interface. The direction of the step formed around the barrier dramatically changes the overall current through the device. In FIG.\ref{asy-i-flat}(a) the potential $V_{a}$ causes spikes in the current up to the corresponding current in FIG.\ref{step-i-flat}(a). If the potentials in regions $a$ and $c$ are reversed the overall current will rise corresponding to the positive side of FIG.\ref{step-i-flat}(a). This effect is present in the subsiquent plots in FIG.\ref{asy-i-flat}; the step created between $V_{a}$ and $V_{c}$ creates similar properties for the other plots; but at a shifted current.
	The conductance plots in FIG.\ref{asy-g-flat} then show very similar properties as FIG.\ref{step-g-flat}; the $|E|$ term causes linear conductance. The exception with the three region devices there is an extra region of low conductance at the barrier height $V_{b}$.

%%%%%%%%%%
%%%%%%%%%%
%%%%%%%%%%
%%%%%%%%%%
%%%%%%%%%%

\subsection{Comparison With Experimental Results}
	The purpose of a theoretical model is to either predict the properties of a material, or to identify the properties of a sample once experimental results are obtained. The theoretical model here is derived for charge carriers in infinite sheet graphene near a Dirac point, however experimentaly this can be difficult to achieve. Instead experimental results for graphene nanoribbons can readily be fabricated via a number of processes \cite{b11, b12,b13,b14}. The results from these nanoribbons can be compared with the theoretical model to determine the properties that the nanoribbons and the infinte sheet share. 
\begin{figure}
	\includegraphics[scale=0.2]{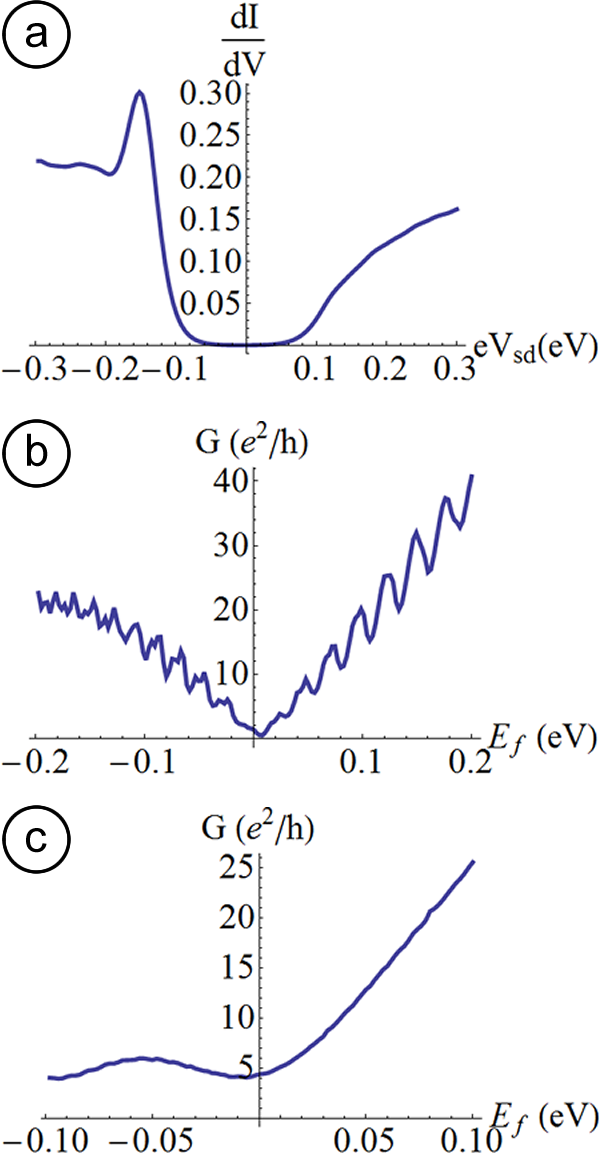}
	\caption{Comparison of theoretical model for infinite graphene sheet against the experimental data for epitaxial graphene nanoribbons \cite{b11}. The theoretical model here uses (b) $T=170$ K, $V_{b}=0.1$ eV, $d=39$ nm, $m_{b}=100$ meV and a shifted $E_{f}$ of $+0.12$ eV. (d) $eV_{sd}=10$ meV, $T=20$ K, $V_{b}=0.9$ eV, $d=100$ nm, $m_{b}=0.6$ eV and an inverted gate dependence on $E_{f}$. (f) $eV_{sd}=70$ meV, $T=55$ K, $V_{b}=0.1$ eV, $d=100$ nm and a shifted $E_{f}$ of $+0.07$ eV.}
	\label{exp-a-flat}
\end{figure}

The results published in \cite{b11} are for epitaxial graphene nanoribbons. The experimental system described here places a graphene nanoribbon bridgeing two large graphene leads grown on silicon carbide. The graphene leads are connected to a current source and voltage probes using a four point contact method. A top gate region between the graphene leads allows the Fermi-energy of the device to be adjusted. 

	The differential conductance in \cite{b11} can be replicated using a theoretical model without the graphene density of states. With the linear dependence removed a barrier with height $V_{b}=0.1$ eV and an energy gap of $m_{b}=0.1$ eV recreates the conductance peaks and zero conductance region. The sample in \cite{b11} shows a sharp change in conductance at $E_{f}=0$, to replicate this the dependence on gate voltage was flipped, then by introducing a high potential barrier, with a large energy gap in the barrier region the asymmetry of the experimental result was simulated. The results displayed from \cite{b11}[supplimentry information] show a smooth dependence of conductance on Fermi level, which implies a small potential was needed to replicate the experimental result, a shift in Fermi level was then required to ensure the minimums in conductance appeared at $E_{f}=0$.

	The use of a large potential barrier with an energy gap agrees with the analysis in \cite{b11}, where it is stated that the asymmery in results is caused by np/pn junctions and that n$\neq$0 subbands experience an energy gap. The theoretical model here does not show a strong temperature dependence, this is possibly due to the experimental results experiencing electronic heating not considered in our model.
\begin{figure}
	\includegraphics[scale=0.2]{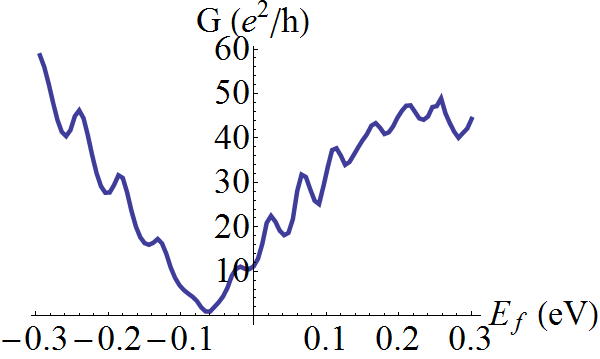}
	\caption{Comparison of theoretical model for infinite graphene sheet against the experimental data for graphene nanoribbons pattered by plasma etching through a PMMA mask on a graphene flake \cite{b14}. The theoretical model here uses $eV_{sd}=10$ meV, $T=20$ K, $V_{b}=0.7$ eV, $m_{b}=0.2$ eV , $d=60$ nm and a shifted $E_{f}$ of $+0.06$ eV.}
	\label{exp-b-flat}
\end{figure}

	The results in \cite{b14} use graphene nanoribbons pattered by plasma etching through a PMMA (polymethyl methacrylate) mask on a graphene flake, which are contacted via either titanium or gold far from the constriction region in a two probe method. Similarly to FIG. \ref{exp-a-flat} the asymmetry here can be recreated with a high potential barrier, the regular oscillations in the conductance appear with low source-drain voltages and thin barrier regions. The location of the minimum in conductance implies that there is some shift in Fermi level. The oscillations shown in FIG.\ref{exp-b-flat} are caused by Fabry-P\'{e}rot resonances within a single potential barrier, however in \cite{b12} these similar resonances are caused in the graphene between the contacts and the constriction.
\begin{figure}
	\includegraphics[scale=0.2]{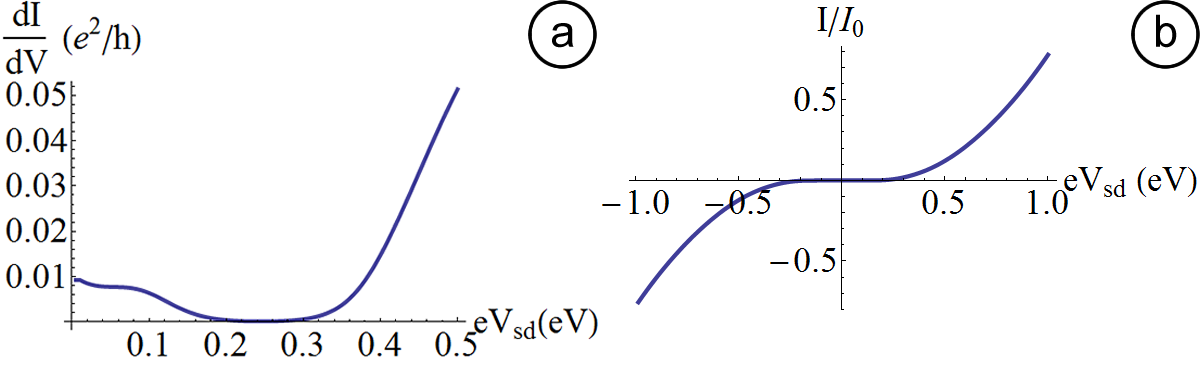}
	\caption{Comparison of theoretical model for infinite graphene sheet against the experimental data for lithographically fabricated graphene nanoribbons \cite{b12}. The theoretical model here uses (b) $T=300$ K, $V_{b}=0.25$ eV, $d=100$ nm and $m_{b}=0.12$ eV. (d) $I_{0}=e2L_{y}/\pi\hbar^{2}v_{f}$, $L_{y}=31$ nm, $T=300$ K, $V_{b}=0$ eV, $d=100$ nm and $m_{b}=0.2$ eV.}
	\label{exp-c-flat}
\end{figure}

	The differential conductance of lithographically fabricated graphene nanoribbons is studied in \cite{b12}. The graphene nanoribbon here is placed on a highly doped silicon substrate with a 285 nm SiO$_{2}$ gate dielectric. These graphene nanoribbons show clear signs of an energy gap. Using the theroretical model these results were replicated with a potential barrier with a height $V_{b}=0.25$ eV and an energy gap $m_{b}=0.12$ eV. The current against source-drain voltage of these samples is shown in FIG.\ref{exp-c-flat}(b), this result resembles charge carriers entering a region with an energy gap $m_{b}=0.2$ eV.

	The analysis in \cite{b12} states that the transport in the considered disordered system is dominated by hopping through localised states. The theoretical model we use with a potential barrier does include localised states and the comparison in FIG.\ref{exp-c-flat} shows there are many similarities in results. Due to the disordered system diffusive transport is present. However, for the nano-scale system considered; there is a high possibiliy that even in the presence of disorder there is some contribution from ballistic transport. The ballistic part of transport may explain the similarities in the shapes of the IV and conductance curves, while the additional disorder in the experimental data allows for the greatly reduced values of current.
\begin{figure}
	\includegraphics[scale=0.2]{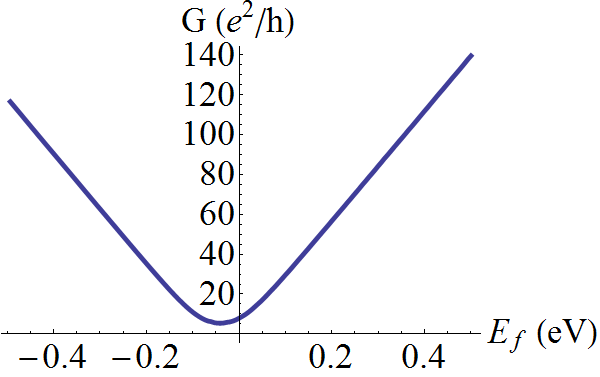}
	\caption{Comparison of theoretical model for infinite graphene sheet against the experimental data for mechanically exfoliated graphene nanoribbons \cite{b13}. The theoretical model here uses $T=300$ K, $eV_{sd}=10$ meV, a Fermi energy shift of $+0.06$ eV and $V_{b}=0.05$ eV.}
	\label{exp-d-flat}
\end{figure}

	In \cite{b13} a graphene nanoribbon fabricated from mechanically exfoliated graphene on p-doped silicon covered with 300 nm thick SiO$_{2}$ is placed between palladium contacts. The mostly linear dependence of conductance on the Fermi energy can be seen in FIG.\ref{exp-d-flat} and implies no scattering region between the two contacts, however at low temperatures the results do show signs of a small scattering region. As the conductance minimum is not at $E_{f}=0$ a shift of $+0.2$ eV has been used in the theoretical model. The anaylsis in \cite{b13} states that the asymmetry of the conductance is likely caused by some form of gate oxide hysteresis. For one sample a gate voltage of 20 V is used, it is stated that this corresponds to a shift in the Fermi level of approximately 260 meV, indicating that there is a large contact resistance present that was not accounted for in the theoretical model.
%%%%%%%%%%
%%%%%%%%%%
%%%%%%%%%%
%%%%%%%%%%
%%%%%%%%%%

\subsection{Conclusion}
	By considering the direction of charge carriers and continuity of probability current as outlined in \cite{b4}, the scattering properties of a graphene potential step were obtained. Using these methods the scattering properties of a Zener barrier; essentialy a barrier on top of a step, were obtained and briefly analysed with respect to properties displayed by potential barriers. Fabry-P\'{e}rot resonances and bound states were both found for the Zener barrier. 
	
	The previously unobtained scattering properties of these Zener tunneling structures were then used to find the current through a nano-device at a finite temperature. A brief derivation of the Landauer formalism \cite{b8} was included to account for the incident angle of incident charge carriers and the graphene density of states \cite{b1}. The IV characteristics were then computed numerically and analysed with respect to gate voltage, temperature and energy gap. Finally the equation for current was reduced to obtain an expression for the conductance of graphene devices.
	
	The IV curves of the graphene step shows similar characteristics to a traditional Zener diode. When an energy gap is included regions of very low current are introduced around low voltages as expected for Zener diodes. The graphene Zener barrier shows similar IV characteristics to the Zener diode, the current obtained is heavily dependant on the direction of the step the barrier is placed upon. When an energy gap is included into the graphene spectrum the graphene transistor shows clear switching capabilities, however the energy gaps required are fairly large.

	With the expressions for current and conductance, the theoretical properties of graphene can be easily compared to those obtained experimentaly. The samples examined from \cite{b11, b12,b13,b14} were replicated, with the possible properties of each sample identified. While many of the experimental results resemble the theoretical model, many features vary by orders of magnitude. The sample bias for nearly all experimental results is vastly greater then the predicted change in Fermi energy, this is possibly due to some form of contact resistance; the voltage applied to the gate region is not perfectly affecting the Fermi level of the graphene. It would therefore require much larger external voltages to change the Fermi level of the sample. The addition of contacts will also change how the current is carried through a sample of graphene; if a voltage probe is placed in an obtrusive manor it may act as an additional scattering region, reducing the flow of current or creating extra resonances between the contacts. The observed effect of temperature is greater then predicted, as described in \cite{b11} an experimental sample will experience some heating when a current is passed through it. It is therefore possible that in order to achieve similar temperature dependences a larger temperature difference will be required. 

	It is clear that due to the many methods that a graphene transistor can be fabricated, further investigation is needed to fully identify the properties of individual samples, this work will hopefully refine the process of creating a switching graphene transistor (see, for a detail the Review \cite{b20}). The inclusion of Zener tunnelling systems will allow a wider range of experimental results to be varified quickly and provide a step towards the development of graphene electronics.

\subsection{Publication}
	This work is an extended version of the work published in \cite{r0}, modified to include the supplementry information and exclude copyrighted figures.
%%%%%%%%%%
%%%%%%%%%%
%%%%%%%%%%
%%%%%%%%%%
%%%%%%%%%%

%%%%%
%%%%%
%%%%%
%%%%%
%%%%%

	\subsection{Supplementary Information}
	Here we present supplementary information for the paper titled "Physical Properties of Zener Tunnelling Nano-devices in Graphene". The information here provides additional derivation for the calculations in the main text, which may be of interest to the reader.
		\subsection{Landauer Formalism in Graphene}
\begin{figure}[h]
				\centerline{\includegraphics[scale=0.5]{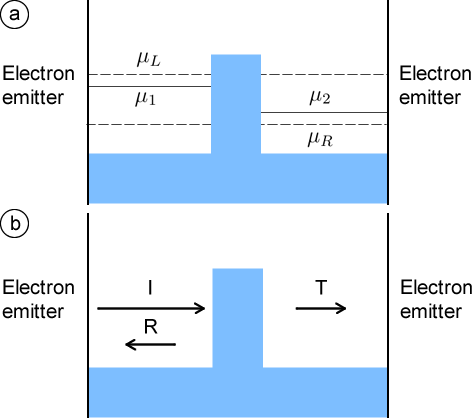}}
				\caption{(Colour Online) (a) Diagram showing quasi-Fermi-energies and chemical potentials of the perfectly conducting wires. Here the left emitter injects electrons up to the quasi-Fermi-energy $\mu_{L}$ and the right emitter injects electrons up the the quasi-Fermi-energy $\mu_{R}$. $\mu_{1}$ and $\mu_{2}$ are the chemical potentials of the perfectly conducting wires to the left and right of the scattering device. (b) A scattering device between two electron emitters. Charge carriers from the left emitter are scattered with a probability $R$ of being reflected and probability $T$ or transmitting through the scattering device.}
				\label{introduction-current}
			\end{figure}

			In this section the Landauer formalism is derived for a graphene scattering device. For a single channel system at non-zero temperatures the current through the system shown in Figure \ref{introduction-current} can be found [28].
			 The system in Figure \ref{introduction-current} consists of 2 incoherent electron reservoirs, which emit charge carriers up to the quasi-Fermi-energy $\mu_{L,R}$, where the subscript $L$ and $R$ represent the reservoir at left or right side of the system respectivly. These reservoirs are then connected to a scattering device via perfect and identical one dimensional conductors. These conductors have chemical potentials $\mu_{1}$ and $\mu_{2}$. The current leaving the left reservoir is then:
			\begin{equation}
				I=ev_{f}\frac{dn}{dE}\left(\mu_{L}-\mu_{R}\right)
				\label{current-dos}
			\end{equation}
			where $e$ is the electron charge, $v_{f}$ is the Fermi velocity and $dn/dE$ is the density of states. The current that is transmitted through the sample is then:
			\begin{equation}
				I=ev_{f}\frac{dn}{dE}T\left(\mu_{R}-\mu_{R}\right)
				\label{current-transmit}
			\end{equation}
			where $T$ is the transmission probability through the scattering device. In [11] the density of states for a single unit cell of graphene at a Dirac point is given by:
			\begin{equation}
				\frac{dn}{dE}=\frac{2A_{c}}{\pi}\frac{|E|}{\hbar^{2}v_{f}^{2}}
				\hspace{1cm}
				A_{c}=\frac{3\sqrt{3}a^{2}}{2}
			\end{equation}
			We define $L_{x}L_{y}/A_{c}$ as the number of unit cells in the sample, where  $L_{x}, L_{y}$ is the size of the sample in the respective dimension. The quantity $L_{x}L_{y}/A_{c}$ shows how many graphene unit cells are present in our sample. As only the $x$-direction current will be considered here, the current in the $x$-direction will be the same in each cell, therefore only the number of graphene unit cells in the $y$-direction will affect the $x$-directional current. This way the quantity $L_{x}$ can be set to one and removed from the calculation. The current through the graphene sample from equation (\ref{current-transmit}) in the $x$-direction becomes:
			\begin{equation}
				I_{x}=e\frac{2L_{y}}{\pi \hbar^{2}v_{f}}T\left(E,\theta\right)\left(\mu_{L}-\mu_{R}\right)|E|cos\left(\theta\right)
			\end{equation}
			The energy and theta dependence for $T$ has been included here to allow for the graphene transmission probability. At non-zero temperatures the states are instead filled according to the corresponding Fermi-Dirac distribution.
			\begin{equation}
				f\left(E-\mu_{L,R}\right)=\frac{1}{e^{\frac{E-\mu_{L,R}}{k_{b}t}}+1}
			\end{equation}
			The current must then be integrated over all energies to account for all states in the Fermi-Dirac distributions.
			\begin{widetext}
			\begin{equation}
				I_{x}=I_{0}\int^{\infty}_{-\infty}\int^{\pi/2}_{-\pi/2}T\left(E,\theta\right)\left[f\left(E-\mu_{L}\right)-f\left(E-\mu_{R}\right)\right]|E|cos\left(\theta\right)dEd\theta
			\end{equation}
			\end{widetext}
			with the constant $I_{0}=e\frac{2L_{y}}{\pi\hbar^{2}v_{f}}$. This result for current can then be used with the definition of conductance, $G=I/V$ to find the conductance at a finite temperature for graphene. The potential difference $V$ is determined by the number of charges on the left and right of the scattering device. This can be found by considering the chemical potentials of the perfectly conducing wires. The chemical potentials $\mu_{1,2}$ must be between the quasi-Fermi-energies of the electron emitters $\mu_{L,R}$. The positioning of these chemical potentials requires that the number of occupied states (electrons) above $\mu_{1}$ is equal to the number of unoccupied states (holes) below $\mu_{1}$, and likewise for states above and below $\mu_{2}$. As all states below $\mu_{R}$ must be filled, only the energy range between $\mu_{L}$ and $\mu_{R}$ needs to be considered. Allowing for positive and negitive velocities the number of states between this range is $2\left(dn/dE\right)\left(\mu_{L}-\mu_{R}\right)$. To the right of the scattering device the number of occupied states is the total number of states available in the wire multiplied by the transmission probability; $T\left(dn/dE\right)\left(\mu_{L}-\mu_{2}\right)$. The number of unoccupied states must therefore be the total number of states available in the wire minus the filled states $\left(2-T\right)\left(dn/dE\right)\left(\mu_{2}-\mu_{R}\right)$. As the number of occupied states is equal to the number of unoccupied states we can write:
			\begin{equation}
				T\left(dn/dE\right)\left(\mu_{L}-\mu_{2}\right)=\left(2-T\right)\left(dn/dE\right)\left(\mu_{2}-\mu_{R}\right)
				\label{mu-2}
			\end{equation}
			On the left of the scattering device the number of occupied states includes those filled by incident and reflected charge carriers $\left(1+R\right)\left(dn/dE\right)\left(\mu_{L}-\mu_{1}\right)$. The number of unoccupied states is then $\left(2-\left(1+R\right)\right)\left(dn/dE\right)\left(\mu_{1}-\mu_{R}\right)$. The number of occupied and unoccupied states must be equal, therefore:
			\begin{align}
				\left(1+R\right)\left(dn/dE\right)&\left(\mu_{L}-\mu_{1}\right)=\\
				&\left(2-\left(1+R\right)\right)\left(dn/dE\right)\left(\mu_{1}-\mu_{R}\right)
				\label{mu-1}
			\end{align}
			The potential difference between the two wires caused by the scattering device is then:
			\begin{equation}
				eV=\mu_{1}-\mu_{2}
			\end{equation}
			Using equations (\ref{mu-2}) and (\ref{mu-1}) the potential difference across the sample is then:
			\begin{equation}
				eV=R\left(\mu_{L}-\mu_{R}\right)
			\end{equation}
			However, at non-zero temperatures the electron emitters fill the states according to the Fermi-Dirac distibutions. To determine the potential difference at non-zero temperatures equations (\ref{mu-2}) and (\ref{mu-1}) can be multipled by the available energy range according to the Fermi-Dirac distributions. Here we will define:
			\begin{equation}
				\frac{-df}{dE}=\left[f\left(E-\mu_{L}\right)-f\left(E-\mu_{R}\right)\right]/\left(\mu_{L}-\mu_{R}\right)
			\end{equation}
			and integrate with respect to energy. This produces the potential difference at non-zero temperatures:
			\begin{equation}
				eV=\frac{\int R\left(E,\theta\right) \frac{-df}{dE} \frac{dn}{dE}dE}{\int\frac{-df}{dE} \frac{dn}{dE}dE}\left(\mu_{L}-\mu_{R}\right)
			\end{equation}
			Using this expression for the voltage and the definition of conductance $G=I/V$ the conductance through a scattering device in graphene can be written as:
			\begin{widetext}
			\begin{equation}
				G_{x}=e^{2}\frac{2L_{y}}{\pi\hbar^{2}v_{f}}\int^{\infty}_{-\infty}\int^{\pi/2}_{-\pi/2}T\left(E,\theta\right)\left[f\left(E-\mu_{L}\right)-f\left(E-\mu_{R}\right)\right]|E|cos\left(\theta\right)dEd\theta\times\frac{\int\frac{-df}{dE} \frac{dn}{dE}dE}{\int R\left(E,\theta\right) \frac{-df}{dE} \frac{dn}{dE}dE}\frac{1}{\left(\mu_{L}-\mu_{R}\right)}
			\end{equation}
			\end{widetext}
			However, the transmission probability in graphene will become one under resonance conditions, Klein tunneling or if $\theta=0$. This will cause the reflection probability $R$ to become zero and the voltage to become zero. To allow for this reference [31] states that any electric field can be absorbed by a finite region of perfect conductor causing the reflection probability over the entire sytem to become one. This effect may also be caused by introducing many scattering devices [28] such as measurement probes. Using these methods the reflection $R \approx 1$ and the one-dimensional conductance for large graphene systems will reduce to:
			\begin{widetext}
			\begin{equation}
				G_{x}=e^{2}\frac{2L_{y}}{\pi\hbar^{2}v_{f}}\int^{\pi/2}_{-\pi/2} \int^{\infty}_{-\infty} T\left(E, \theta\right)\frac{f\left(E-\mu_{L}\right)-f\left(E-\mu_{R}\right)}{\mu_{L}-\mu_{R}}|E|cos\theta dE d\theta
			\end{equation}
			\end{widetext}
			At zero temperature and for small voltages, the Fermi distributions become the Dirac delta function centered at the Fermi energy $E_{f}$. With the identity $\int f(x)\delta(x) dx=f(0)$ the zero temperature conductance for small voltages and large systems becomes:
			\begin{equation}
				G_{x}= G_{0}\int^{\frac{\pi}{2}}_{-\frac{\pi}{2}} T\left(E_{f},\theta\right)|E_{f}|cos\left(\theta\right)d\theta\\
			\end{equation}
			where $G_{0}=e^{2}\frac{2L_{y}}{\pi\hbar^{2}v_{f}}$.
\end{document}